# Experimental efforts in search of $^{76}$Ge Neutrinoless Double Beta Decay

**Somnath Choudhury** †

*Department of Physics & Meteorology*
*Indian Institute of Technology, Kharagpur – 721302, India*

**Abstract**

Neutrinoless double beta decay is one of the most sensitive approaches in non-accelerator particle physics to take us into a regime of physics beyond the standard model. This article is a brief review of the experiments in search of neutrinoless double beta decay from $^{76}$Ge. Following a brief introduction of the process of double beta decay from $^{76}$Ge, the results of the very first experiments IGEX and Heidelberg-Moscow which give indications of the existence of possible neutrinoless double beta decay mode has been reviewed. Then ongoing efforts to substantiate the early findings are presented and the Majorana experiment as a future experimental approach which will allow a very detailed study of the 0ν decay mode is discussed.

## 1. Introduction

Neutrinoless double beta decay is one of the most sensitive approaches with great perspectives to test particle physics beyond the Standard Model. There is immense scope to use 0νββ decay for constraining neutrino masses, left–right–symmetric models, interactions involving R-parity breaking in the supersymmetric model and leptoquark scenarios, as well as effective lepton number violating couplings. Experimental limits on 0νββ decay are not only complementary to accelerator experiments but at least in some cases competitive or superior to the best existing direct search limits. The steadily improving experimental limits on the half-life of 0νββ can be translated into more stringent limits on the parameters of these new physics scenarios.

In the process of beta decay an unstable nucleus decays by converting a neutron in the nucleus to a proton and emitting an electron and an anti-neutrino. In order for beta decay to be possible the final nucleus must have a larger binding energy than the original nucleus. For some nuclei, such as Germanium-76 the nuclei with atomic number one higher have a smaller binding energy, preventing beta decay from occurring. In the case of Germanium-76 the nuclei with atomic number two higher, Selenium-76 has a larger binding energy, so the "double beta decay" process is allowed. In double beta decay two neutrons in the nuclei are converted to protons, and two electrons and two anti-neutrinos are emitted. It is the rarest known kind of radioactive decay; it was observed for only ten isotopes. For some nuclei, the process occurs as conversion of two protons to neutrons, with emission of two neutrinos and absorption of two orbital electrons (double electron capture). If mass difference between the parent and daughter atoms is more than 1022 keV (two electron masses), another branch of the process becomes possible, with capture of one orbital electron and emission of one positron. And, at last, when the mass difference is more then 2044 keV (four electron masses), the third branch of the decay arises, with emission of two positrons ($\beta^+\beta^+$ decay).

The processes described above are also known as two neutrino double beta decay, as two neutrinos (or anti-neutrinos) are emitted. If the neutrino is a Majorana particle, meaning that the anti-neutrino and the neutrino are actually the same particle then it is possible for neutrinoless double beta decay to occur. In 0νββ decay the emitted neutrino is immediately absorbed (as its anti-particle) by another nucleon of the nucleus, so the total kinetic energy of the two electrons would be exactly the difference in binding energy between the initial and final state nuclei.

Experiments have been carried out and proposed to search for 0νββ decay mode, as its discovery would indicate that neutrinos are indeed Majorana particles and allow a calculation of neutrino mass. While the two-neutrino mode (1.1) is allowed by the Standard Model of particle physics, the neutrinoless mode (0νββ) (1.2) requires violation of lepton number (ΔL=2). This mode is possible only, if the neutrino is a Majorana particle, i.e. the neutrino is its own antiparticle. Double beta decay, the rarest known nuclear decay process, can occur in different modes:

2νββ -decay : A(Z,N) → A(Z+2, N-2)+2e⁻ +2 $\bar{\nu}$ (1.1)
0 ν ββ -decay : A(Z,N) → A(Z+2, N-2) + 2e⁻ (1.2)
0 ν(2) χ ββ -decay : A(Z,N) → A(Z+2, N-2)+2e⁻ + (2) χ (1.3)

## 2. Double Beta Decay: A Rare Process

The process arises in certain cases of even-A nuclei, where A is the mass number and is the sum of the number of protons and neutrons (A = Z + N). For even-A nuclei, the strong pairing force between like nucleons (neutrons like to be paired with other neutrons in a given nucleus, with the same true for protons), the binding energy of even-even nuclei (even number of protons and even number neutrons) is larger than that of odd-odd nuclei (odd numbers of protons and neutrons). This fact results in two separate parabolas on a plot of binding energy, one parabola for even-even nuclei and one for odd-odd. Consequently, one occasionally finds a situation where two even-even nuclei for a given mass number A are stable against ordinary beta decay. However, the heavier nucleus is not fully stable and can decay to the lighter nucleus via normal double beta decay, a second-order process whereby the nuclear charge changes by two units. The ground state of the even-even nuclei is 0⁺ (positive parity) and the nuclear transition is 0⁺→0⁺.

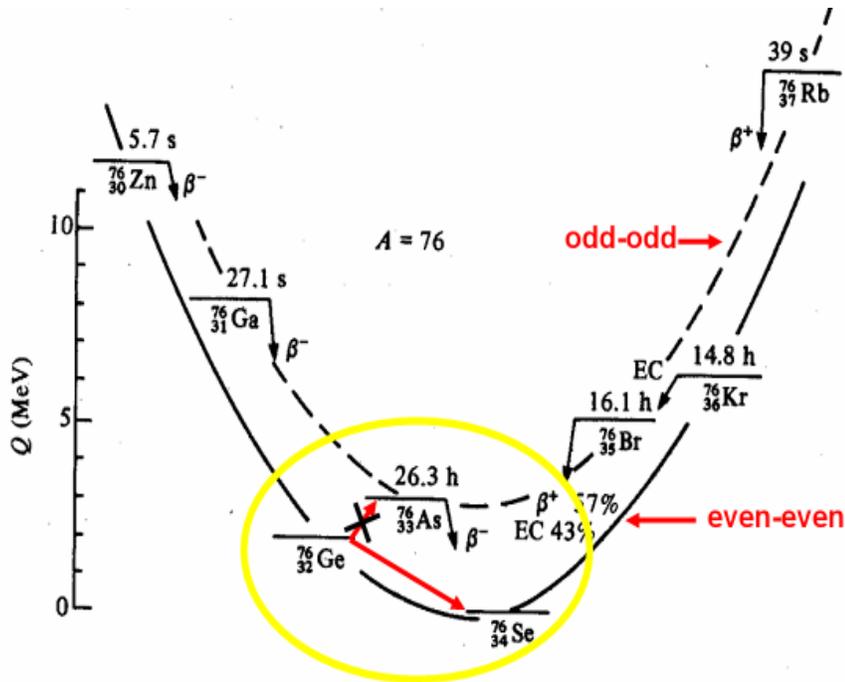

Ground states of even-even nuclei: 0⁺

One particular type of experimental approach that hopes to determine if the neutrino is a massive Majorana particle is the search for neutrinoless double beta decay. This type of experiment is perhaps the only feasible method for determining if the neutrino is a Majorana or Dirac particle. While neutrinoless double beta decay has not yet been experimentally discovered, searches have been conducted for many years, with many continuing today. In fact, the next generation of double beta decay experiments is currently being designed and developed and involves a tremendous increase in the amount of source material to be studied (on the order of a half-ton or more). In neutrinoless double beta decay an antineutrino emitted at the first vertex is absorbed at the second as seen in the figure below or that a virtual neutrino emitted by a neutron is absorbed by the second neutron participating in the double beta decay.

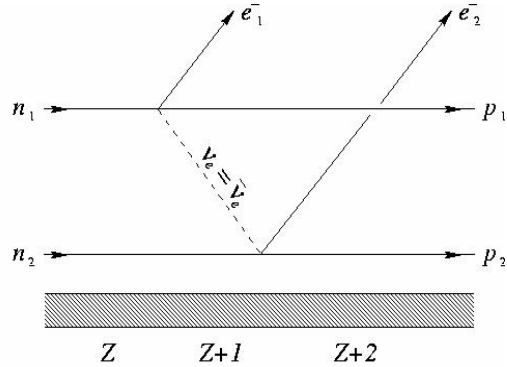

The two neutrino mode is allowed in standard model. The neutrinoless mode can occur only if neutrinos have masses of the Majorana type. The decay rate is proportional to the squared mass. In other words, the half life is inversely proportional to the squared mass. Experimentally one can distinguish the two modes. In the two neutrino mode the electrons take away only a fraction of the energy Q released in the decay. The sum energy spectrum is continuous, extending from 0 to Q. In the neutrinoless mode the total energy Q is carried away by the electrons, and the sum energy spectrum is a peak centered at Q, with a width given by the instrumental resolution.

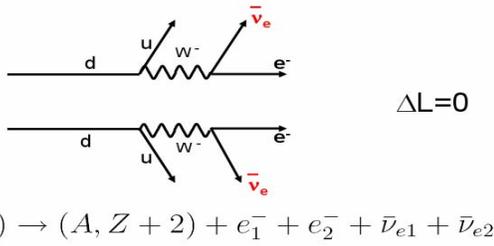

$(A, Z) \rightarrow (A, Z+2) + e_1^- + e_2^- + \bar{\nu}_{e1} + \bar{\nu}_{e2}$

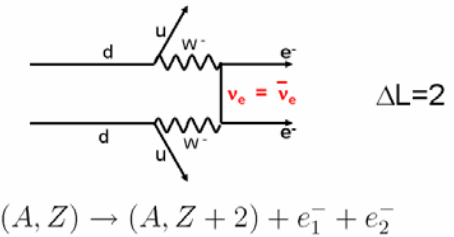

$(A, Z) \rightarrow (A, Z+2) + e_1^- + e_2^-$

The decay rate for 2ν ββ decay which is allowed in the Standard Model of physics is given by

$$\left[T_{1/2}^{2\nu}(0^+ \to 0^+)\right]^{-1} = G^{2\nu}(E_0, Z) \left| M_{GT}^{2\nu} - \frac{g_V^2}{g_A^2} M_F^{2\nu} \right|^2 \tag{2.1}$$

The decay rate for the process involving the exchange of the Majorana neutrino in the absence of right-handed currents can be expressed as follows:

$$\left[T_{1/2}^{0\nu}(0^+ \to 0^+)\right]^{-1} = G^{0\nu}(E_0, Z) \left| M_{GT}^{0\nu} - \frac{g_V^2}{g_A^2} M_F^{0\nu} \right|^2 <m_\nu>^2 \tag{2.2}$$

The $M_{GT}$ and $M_F$ are the nuclear matrix elements of Gamow-Teller and Fermi transitions respectively. The nuclear matrix elements of the $0^+ \to 0^+$ Gamow-Teller and Fermi transition for the two neutrino mode in weak theory in the second order perturbation is given by

$$M_F^{2\nu} = \sum_n \frac{\langle 0_f^+ | \sum_j \tau_j^+ | 1_n^+ \rangle \langle 1_n^+ | \sum_k \tau_k^+ | 0_i^+ \rangle}{E_n - E_i + \Delta} \tag{2.3}$$

$$M_F^{2\nu} = \sum_n \frac{\langle 0_f^+ | \sum_j \vec{\sigma}_j \tau_j^+ | 1_n^+ \rangle \langle 1_n^+ | \sum_k \vec{\sigma}_k \tau_k^+ | 0_i^+ \rangle}{E_n - E_i + \Delta} \qquad (2.4)$$

where $\Delta$ denotes the average energy and is given by $\Delta = (E_i - E_f)/2$ and the Gamow-Teller transition operator and the Fermi transition operator is given by $\sum_j \sigma_j \tau_j^+$ and $\sum_j \tau_j^+$ respectively.

A complete orthonormal set of intermediate excited states have been introduced denoted by $|1_n^+\rangle$. Thus the two neutrino double beta decay mode has been expressed in terms of single beta transitions through the introduction of intermediate excited states via which the transition from the initial $0^+$ to the final $0^+$ state occurs.

For the neutrinoless mode, the nuclear matrix elements resulting from Fermi and Gamow-Teller transitions are given by

$$M_F^{0\nu} = \langle 0_f^+ | \sum_{j,k} H(r_{jk}, E) \tau_j^+ \tau_k^+ | 0_i^+ \rangle \qquad (2.5)$$

$$M_{GT}^{0\nu} = \langle 0_f^+ | \sum_{j,k} H(r_{jk}, E) \vec{\sigma}_\mathbf{j} \cdot \vec{\sigma}_\mathbf{k} \, \tau_j^+ \tau_k^+ | 0_i^+ \rangle \qquad (2.6)$$

The function $H$ depends on the distance between the nucleons and approximately has the form

$$H(r, E) = \frac{2R}{\pi r} \int_0^\infty dq \frac{q \sin qr}{\omega \{\omega + E - (E_i + E_f)/2\}} \qquad (2.7)$$

where, $R = r_o A^{1/3}$, A being the mass number and $r_o = 1.2$ fm.

The part $G^{2\nu}$ and $G^{0\nu}$ results from integrating over the lepton phase space, and $g_V$ and $g_A$ are the weak vector and axial-vector coupling constants respectively. The $<m_\nu>$ is the effective electron neutrino mass. If the light neutrino ($m_j \ll$ few MeV) exchange is the dominant mechanism for the 0νββ-decay process and that both the neutrino currents are left-handed, then the 0νββ-decay amplitude is proportional to the lepton number violating parameters. This effective mass is related to the light neutrino mass eigenvalues ($m_j$) and the mixing parameters ($U_{ej}$) and is give by the relation

$$<m_\nu> = \left| \sum_j m_j U_{ej}^2 \right| \qquad (2.8)$$

The effective light neutrino mass $<m_\nu>$ may be suppressed by a destructive interference between the different contributions in the sum of equation (2.8) if CP is conserved. In this case the mixing matrix satisfies the condition $U_{ej} = U_{ej}^* . \zeta_j$, where $\zeta_j = \pm i$ is the CP parity of the Majorana neutrino $\nu_j$. The absolute value has thus been inserted for convenience, since the quantity inside it is squared in equation (2.8) and is complex if CP is violated.

The ideal 0νββ-decay experiment has the following dream features: the lowest possible background, the best possible energy resolution, the greatest possible mass of the parent isotope, detection efficiency near 100% for valid events, a unique signature and the lowest possible construction cost. The next sections review experiments in such an effort from the isotope $^{76}$Ge.

## 3. International Germanium EXperiment (IGEX)

The nuclear Double Beta Decay is a unique ground to investigate the nature and properties of the neutrino. The neutrinoless decay mode, if it exists, would provide an unambiguous evidence of the Majorana nature of the neutrino, its non-zero mass, and the non-conservation of lepton number. After implication from solar and atmospheric neutrino oscillation results that neutrinos have non-zero mass, the process of neutrinoless Double Beta Decay has become the most relevant place to test the neutrino mass scale and its hierarchy pattern. To achieve high sensitivity limits of the effective Majorana electron neutrino mass derived from the neutrinoless half-life lower bound required for such new objectives, it will require a large number of double beta emitter nuclei, a very low background and a sharp energy resolution in the Q-value region, and effective methods to disentangle signal from noise. A typical example of this type of search was the IGEX. The International Germanium EXperiment (IGEX) was a search for the neutrinoless double beta decay of $^{76}$Ge employing large amounts of HPGe detectors, isotopically enriched to 86% in $^{76}$Ge. In the first phase of the experiment three detectors of 0.7 kg active volume each were operated: one in the Homestake gold mine (4000 m.w.e.), other in the Baksan Neutrino Observatory (660 m.w.e.) and the other in the Canfranc underground laboratory (Laboratory 2 at 1380 m.w.e.). A conservative lower bound on the neutrinoless half-life of about $10^{24}$ years was derived.

The International Germanium EXperiment (IGEX) took data at the Canfranc Underground Laboratory in Spain at a depth of 2450 m.w.e. in a search of neutrinoless double beta decay. Three Germanium detectors (RG1, RG2 and RG3), of ~2 kg each, enriched to 86% in $^{76}$Ge were used. Efforts were made to reduce part of the radioactive background by discriminating it from the expected signal by comparison of the shape of the pulses (PSD) of both types of events. The method was applied to the data recorded by two Ge detectors of the IGEX, which has produced one of the two best current sensitivity limits for the Majorana neutrino mass parameter. In the second phase, three large detectors (2 kg each) were fabricated (with improvements derived from the analysis of data of Phase 1). They are installed in the Canfranc underground laboratory (Laboratory 3 at 2450 m.w.e.) inside a low background shielding consisting of 40 cm of lead, a PVC box (silicone sealed and flushed with nitrogen), 2mm of cadmium, 20 cm of polyethylene and an active veto (plastic scintillators). A pulse shape discrimination (PSD) technique capable to distinguish single site events (ββ decay events for example) from multisite events (the most dominant background events) is implemented. New limits on the neutrinoless half-life and the neutrino mass parameter were thus obtained from here.

In large intrinsic Ge detectors, the charge carriers take 300 - 500 ns to reach their respective electrodes. These drift times are long enough for the current pulses to be recorded at a sufficient sampling rate. The current pulse contributions from electrons and holes are displacement currents, and therefore dependent on their instantaneous velocities and locations. Accordingly, events occurring at a single site (ββ-decay events for example) have associated current pulse characteristics which reflect the position in the crystal where the event occurred. More importantly, these single-site events (SSE) frequently have pulse shapes that differ significantly from those due to the background events that produce electron-hole pairs at several sites by multi-Compton-scattering process, for example (the so-called Multi-Site Events (MSE)). Consequently, pulse-shape analysis was used to distinguish between these two types of energy depositions since DBD events belong to the SSE class of events and will deposit energy at a single site in the detector while most of the background events belong to the MSE class of events and will deposit energy at several sites. It provided a rejection of ~ 60 % of the events in the region of interest, accepting the criterion that those events having more than two lobes cannot be due to DBD event.

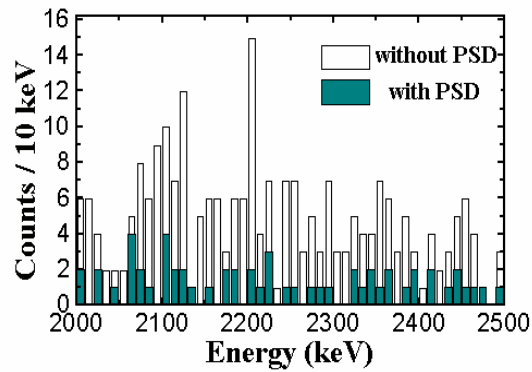

IGEX spectrum with and without the PSD background rejection.

The IGEX detectors had the initial objective of the detection of the double beta decay of $^{76}$Ge. At the end of 1999 certain modifications were made to adapt the detectors to the detection at low energy where the signal of WIMPs (Weak Interacting Massive Particles) is relevant. The shielding, shared by three IGEX detectors (2 kg germanium detectors isotopically enriched to 86% in $^{76}$Ge) and the COSME detector, included from inside to outside 40 cm of lead, a PVC box (silicone sealed and flushed with nitrogen), 2 mm of cadmium, plastic scintillators working in anticoincidence with the Ge detectors and 20 cm of polyethylene. The shielding was modified on July 2001 as it included only one 2 kg germanium detector inside a more efficient neutron shielding. These techniques of passive and active shielding, along with the extreme radiopurity of the detectors and their components, allowed a low energy background as well as a low enough threshold which are unique in this type of detectors. So, very stringent contour limits for cross sections and masses of dark matter particles interacting with Ge nuclei through spin-independent interactions were derived from here. The need to understand and reject backgrounds in Ge-diode detector double-beta decay experiments thus gave rise to the development of the pulse shape analysis technique in such detectors to distinguish DBD single-site energy deposits from the multiple-site deposits. Henceforth the analysis was extended by DBD people to segmented Ge detectors to study the effectiveness of combining segmentation with pulse shape analysis to identify the multiplicity of the energy deposits.

The IGEX calculations for a lower bound to the half-life for the neutrinoless mode where there were fewer than 3.1 candidate events (90% Confidence Level) under a peak having FWHM = 4 keV and centered at 2038.56 keV corresponded to:

$$T_{1/2}^{0\nu}(^{76}Ge) > \frac{4.87 \times 10^{25}}{3.1} yr \cong 1.57 \times 10^{25} yr$$

The requirements for a next generation experiment can easily be deduced by reference to $T_{1/2}^{0\nu} = \frac{(\ln 2).Nt}{c}$ (3.1) where N is the number of parent nuclei, t is the counting time, and c is the upper limit on the number of 0νββ-decay counts consistent with the observed background. To improve the sensitivity of ‹$m_\nu$› by a factor of 100, the quantity Nt/c must be increased by a factor of $10^4$. The quantity N can feasibly be increased by a factor of $\sim 10^2$ over present experiments, so that t/c must also be improved by that amount. Since practical counting times can only be increased by a factor of 2 to 4, the background should be reduced by a factor of 25 to 50 below present levels. These are approximately the target parameters of the next generation neutrinoless double-beta decay experiments.

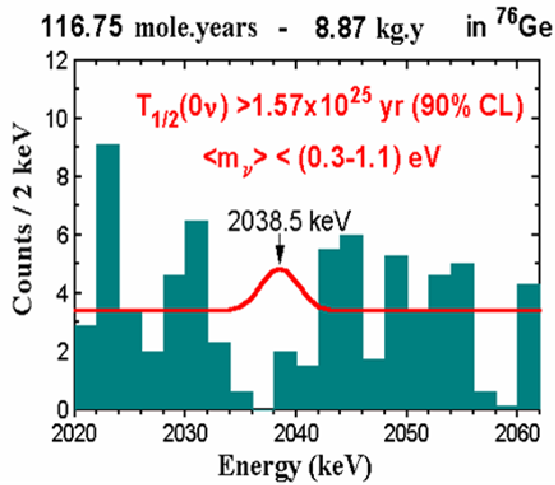

Histogram of the IGEX data in the energy region of interest for the 0ν -ββ decay. The limits on the half-life and neutrino mass parameter are also shown.

The Effective ν Mass: The section of KKDK on effective neutrino mass ("Critical View to the IGEX neutrinoless double-beta decay experiment..." published in Phys. Rev. D, Volume 65 (2002) 092007, by H. V. Klapdor-Kleingrothaus, A. Dietz, and I. V. Krivosheina) begins with: "Starting from their incorrectly determined half-life limit the authors claim a range of effective neutrino mass of (0.33-1.35) eV." In response the IGEX collaboration, came out stating that KKDK selected only the 52.51 mole·years of the IGEX data that had been subjected to PSD and obtained $T_{1/2}^{0\nu} > 7.1 \times 10^{24}$ y using the maximum number of counts, 3.1, from the entire 117 mole·years of data which was erroneous and unjustified. In another case, KKDK also decided to arbitrarily use the entire IGEX data set prior to PSD selection from which they obtained 0ν a bound of $T_{1/2}^{0\nu} > 1.1 \times 10^{25}$ y for which there was no scientific justification for selecting only PSD corrected data on one hand and totally ignoring the PSD corrected data on the other hand. In the conclusion of KKDK it states: "the IGEX paper - apart from the too high half-life limits presented, as a consequence of an arithmetic error - is rather incomplete in its presentation". In response to this paper the IGEX collaboration published the article "The IGEX experiment revisited: a response to the critique of Klapdor-Kleingrothaus, Dietz, and Krivosheina" where they stated that there was absolutely no arithmetic error and that the analysis of the published IGEX data presented in KKDK stands illegitimate. To obtain a much shorter bound on the half-life, they arbitrarily analyzed two ~ halves of the data separately. Instead of having $4.88 \times 10^{25}$ y in the numerator (ln2 N.t) they used $2.2 \times 10^{25}$ y. Yet they used the 90% CL upper limit on the number of counts under the peak, obtained by IGEX from all of the data. In another analysis, they ignored the fact that 52.51 mole·years were corrected with PSD and treated the complete uncorrected data set. Naturally, the lower limits on $T_{1/2}^{0v}$ ($^{76}$Ge) obtained by these completely unjustified procedures are shorter than that obtained from properly analyzing the complete data set. This paper henceforth states "the lower limit quoted by IGEX, $T_{1/2}^{0v} \geq 1.57 \times 10^{25}$ years, is correct and that there was no arithmetical error as claimed in the Critical View article."

## 4. The HEIDELBERG - MOSCOW Experiment

The Heidelberg-Moscow experiment at the Gran Sasso underground laboratory is now claimed to be the most sensitive neutrinoless double beta decay experiment worldwide. It has contributed in an extraordinary way to the research in neutrino physics and particularly to beyond standard model physics, and limits for the latter are competing with those from the largest high-energy accelerators. The emphasis on the first indication for neutrinoless double beta decay is found in the Heidelberg-Moscow experiment giving first evidence of the lepton number violation and a Majorana nature of the neutrinos. The neutrinoless double beta decay could answer questions to the absolute scale of the neutrino mass and the fundamental character of the neutrino whether it is a Dirac or a Majorana particle.

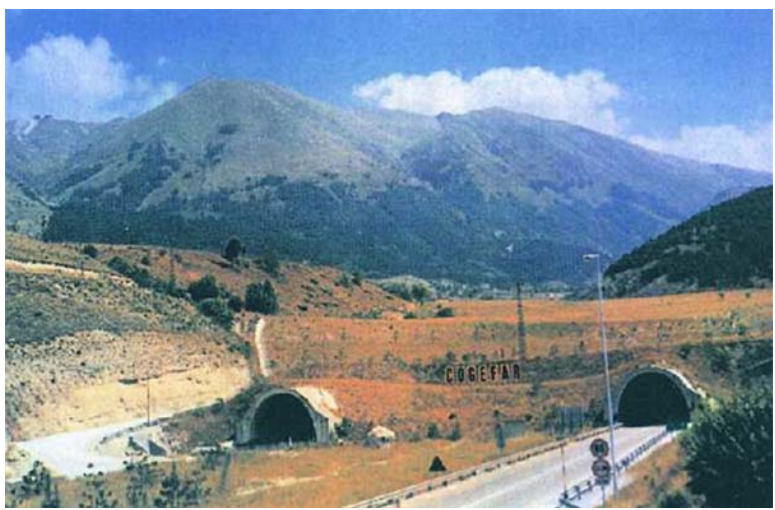

Entrance of the highway tunnel under Gran Sasso mountain.

With the support of the LNGS the experimental building of the experiment was built between Halls A and B in Gran Sasso, into which the first enriched $^{76}$Ge detector (the first high-purity enriched $^{76}$Ge detector worldwide) was installed in July 1990 . First preparation work had been done since 1989 in a provisional tent in Hall C. The full amount of five enriched $^{76}$Ge detectors of in total 11 kg was finally installed in 1995 and were operated since 1996 with a newly developed pulse shape discrimination method

High purity germanium crystals, enriched by Germanium-76 isotope up to 86% are used as the main detecting elements. Five coaxial detectors with the total weight of 11.5 kg (125 moles in the active volume of detectors) are used. Each detector is located in a separate cryostat made of electrolytic copper with low content of radioactive impurities. The quantity of other designed materials (iron, bronze, light material insulators) is minimized in order to reduce the feasible radioactive impurities contribution to the total background of the detectors. The detectors were located in two separate shielded boxes. One of them, 270 mm thick is made of electrolytic copper (detector #4), the other consists of two layers of lead – inner -100mm of high purity LCD2-grade lead and outer – 200 mm of low background Boliden lead (detectors ## 1,2,3,5). Each setup is coated with stainless steal casing. Non-radioactive pure nitrogen was blown through casings to reduce radon emanation contribution. To reduce neutron background the casing with detectors ##1,2,3,5 was coated with borated polyethylene and two anticoincidence plates of plastic scintillator were located over the casing in order to reduce muon component. The setup was located in Gran Sasso underground laboratory, Italy at a depth of 3500 metres of water equivalent of the lab reduces influence of cosmic rays on background conditions of the experiment. The electronics and the system of collecting data allow to record each event – the number (or numbers) of acted detector, amplitude and pulse shape, and anticoincidence veto. The Heidelberg-Moscow experiment, with five enriched 86%-88% high-purity p-type Germanium detectors, of in total 10.96 kg of active volume, used the largest source strength of all double beta experiments at present, and reached a record low level of background. The detectors were the first high-purity Ge detectors ever produced. The degree of enrichment was checked by investigation of tiny pieces of Ge after crystal production using the Heidelberg MP-Tandem accelerator as a mass spectrometer.

The detectors, except detector # 4, were operated in a common Pb shielding of 30 cm, which consisted of an inner shielding of 10 cm radiopure LC2-grade Pb followed by 20 cm of Boliden lead. The whole setup was placed in an air-tight steel box and flushed with radiopure nitrogen in order to suppress the $^{222}$Rn contamination of the air. The shielding was improved in the course of the measurement. The steel box operated since 1994 centered inside a 10-cm boron-loaded polyethylene shielding to decrease the neutron flux from outside. An active anticoincidence shielding was placed on top of the setup since 1995 to reduce the effect of muons. Detector # 4 was installed in a separate setup, which had an inner shielding of 27.5 cm electrolytical Cu, 20 cm lead, and boron-loaded polyethylene shielding below the steel box, but no muon shielding. The setup was kept air-tight closed since installation of detector #5 in February'95. Since then no radioactive contaminations of the inner of the experimental setup by air and dust from the tunnel could occur.

The sensitivity for the 0ν ββ half-life is given by $\quad T^{0\nu}_{1/2} \sim a \times \varepsilon \sqrt{\dfrac{M.t}{\Delta E.B}} \quad (and \quad \dfrac{1}{\sqrt{T^{0\nu}}} \sim \langle m_\nu \rangle)$ (4.1)

With $a$ denoting the degree of enrichment, ε the efficiency of the detector for detection of a double beta event, M the detector (source) mass, ΔE the energy resolution, B the background and t the measuring time, the sensitivity of our 11 kg of enriched $^{76}$Ge experiment corresponds to that of an at least 1.2 ton natural Ge experiment. After enrichment - the other most important parameters of a ββ experiment are: energy resolution, background and source strength. The high energy resolution of the Ge detectors of 0.2% or better, assures no background for a 0νββ line from the two-neutrino double beta decay in this experiment ($5.5 \times 10^{-9}$ events expected in the energy range 2035-2039.1keV), in contrast to most other present experimental approaches, where limited energy resolution is a severe drawback.

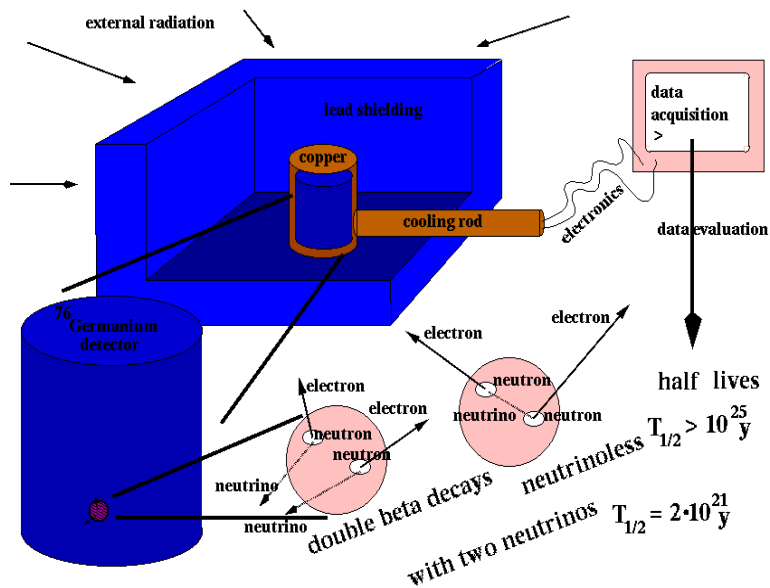

The efficiency of Ge detectors for detection of 0ν ββ decay events is close to 100%. The source strength in the Heidelberg-Moscow experiment of 11kg was the largest source strength ever operated in a double beta decay experiment. The background reached to the experiment, was 0.113 ± 0.007 events/kg y keV (in the period 1995-2003) in the 0ν ββ decay region (around $Q_{\beta\beta}$). This was the lowest limit ever obtained in such type of experiment.

The statistics collected in this experiment during 13 years of stable running is the largest ever collected in a double beta decay experiment. The experiment took data during ~ 80% of its installation time. The Q value for neutrinoless double beta decay was recently determined with high precision.

The background of the experiment: (1) primordial activities of the natural decay chains from $^{238}$U, $^{232}$Th, and $^{40}$K; (2) anthropogenic radio nuclides, like $^{137}$Cs, $^{134}$Cs, $^{125}$Sb, $^{207}$Bi; (3) cosmogenic isotopes, produced by activation due to cosmic rays during production and transport; (4) the bremsstrahlungs spectrum of $^{210}$Bi (daughter of $^{210}$Pb); (5) elastic and inelastic neutron scattering; and (6) direct muon-induced events.

H.V. Klapdor-Kleingrothaus, O. Chkvorez, I.V. Krivosheina and C. Tomei at *Max-Planck-Institut fur Kernphysik* in the Heidelberg-Moscow group presented a paper concerning "Measurement of the $^{214}$Bi spectrum in the energy region around the Q-value of $^{76}$Ge neutrinoless double-beta decay". In this work they presented the measurements of the $^{214}$Bi spectrum from a $^{226}$Ra source with a high purity germanium detector. Their attention was mostly focused on the energy region around the Q-value of $^{76}$Ge neutrinoless double-beta decay (2039.006 keV). The results of the measurement strongly relates to the first indication for neutrinoless double beta decay of $^{76}$Ge. An analysis of the data collected during ten years of measurements by the Heidelberg-Moscow experiment, at Gran-Sasso Underground Laboratory, yields a first indication for the neutrinoless double beta decay of $^{76}$Ge. An important point of this analysis is the interpretation of the background, in the region around the Q-value of the double beta decay (2039.006 keV), as containing several weak photopeaks. It was suggested and has been shown that four of these peaks are produced by a contamination from the isotope $^{214}$Bi, whose lines are present throughout the Heidelberg-Moscow background spectrum.

In this work they performed a measurement of a $^{226}$Ra source with a high-purity germanium detector. The aim of this work was to study the spectral shape of the lines in the energy region from 2000 to 2100keV and, most important, to show the difference in this spectral shape when changing the position of the source with respect to the detector, and to verify the effect of TCS (True Coincidence Summing) for the weak $^{214}$Bi lines seen in the Heidelberg-Moscow experiment. The activity of the $^{226}$Ra source is 95.2kBq. The isotope $^{226}$Ra appears in the $^{238}$U natural decay chain and from its decays also $^{214}$Bi is produced. The γ-spectrum of $^{214}$Bi is clearly visible in the $^{226}$Ra measured spectrum. $^{214}$Bi is a naturally occurring isotope: it is produced in the $^{238}$U natural decay chain through the β- decay of $^{214}$Pb and the alpha decay of $^{218}$At. With a subsequent β- reaction, $^{214}$Bi decays then into $^{214}$Po (the branching ratio with respect to the α decay into $^{210}$Tl is 99.979%). The decay, however, does not lead directly to the ground state of $^{214}$Po, but to its excited states. From the decays of those excited states to the ground state the well known γ-spectrum of $^{214}$Bi is obtained, which contains more than hundred lines.

In the table given below, one can see in the energy region around the Q-value of the 0νββ decay (2000-2100keV), four γ-lines and one E0 transition with energy 2016.7keV are expected. The E0 transition can produce a conversion electron or an electron-positron pair but it could not contribute directly to the γ-spectrum in the considered energy region if the source is located outside the detector active volume.

| Energy (keV) | Intensity(%) |
| --- | --- |
| 2010.71 | 0.050 |
| 2016.7 | 0.0058 |
| 2021.8 | 0.020 |
| 2052.94 | 0.078 |
| 20889.7 | 0.050 |

The intensity of each line is defined as the number of emitted photons, with the corresponding energy, per 100 decays of the parent nuclide. The considerations for the measurement were the efficiency of the detector (which depends on the size of the detector and on the distance source-detector) and the effect called True Coincidence Summing (TCS). The lifetimes of the atomic excited levels are much shorter than the resolving time of the detector. If two gamma-rays are emitted in cascade, there is a certain probability that they will be detected together. If this happens, then a pulse will be recorded which represents the sum of the energies of the two individual photons, instead of two separated pulses with different energies. The TCS effect can result both in lower peak-intensity for full-energy peaks and in bigger peak-intensity for those transitions whose energy can be given by the sum of two lower-energy gamma-rays. In this case, the lines at 2010.7 keV and 2016.7 keV can be given by the coincidence of the 609.312 keV photon (strongest line, intensity = 46.1%) with the 1401.50keV photon (intensity = 1.27%) or with the 1407.98keV photon (intensity = 2.15%). The degree of TCS depends on the probability that two gamma-rays emitted simultaneously will be detected simultaneously which is a function of the detector geometry and of the solid angle subtended at the detector by the source and for this the intensities of the two lines mentioned above (2010.71keV and 2016.7keV) are expected to depend on the position of the source with respect to the detector.

The $^{226}$Ra γ-ray spectra were measured using a γ-ray spectroscopy system based on an HPGe detector installed in the operation room of the HEIDELBERG-MOSCOW experiment in Gran Sasso Underground Laboratory, Italy. The coaxial germanium detector had an external diameter of 5.2cm and 4.9cm height. The distance between the top of the detector and the copper cap was kept at 3.5cm. The relative detection efficiency of the detector was 23% and the energy resolution being 3.6keV for the energy range 2000-2100keV. The measurement of $^{214}$Bi spectrum, with a high purity germanium detector, in the energy region around the Q-value of $^{76}$Ge neutrinoless double-beta decay (2039.006keV) was done with the $^{226}$Ra source used for the measurements positioned, in a first step the source was positioned on the top of the detector, directly in contact with the copper cap (close geometry) and in a second step the source was moved 15cm away from the detector cap (far geometry). The results of the measurements show that, if the source is close to the detector, the intensities of the weak Bi lines in the energy region 2000- 2100keV are not in the same ratio as reported by Table of Isotopes. The results of the analysis of the data collected by the Heidelberg-Moscow experiment with all the five detectors, yielding a first indication for the neutrinoless double beta decay of $^{76}$Ge, shows that four $^{214}$Bi lines are present in the energy region from 2000 to 2080keV (many other strong lines from the same isotope are present in the spectrum), due to the presence of bismuth in the experimental setup, especially in the copper in the vicinity of the Ge crystals.

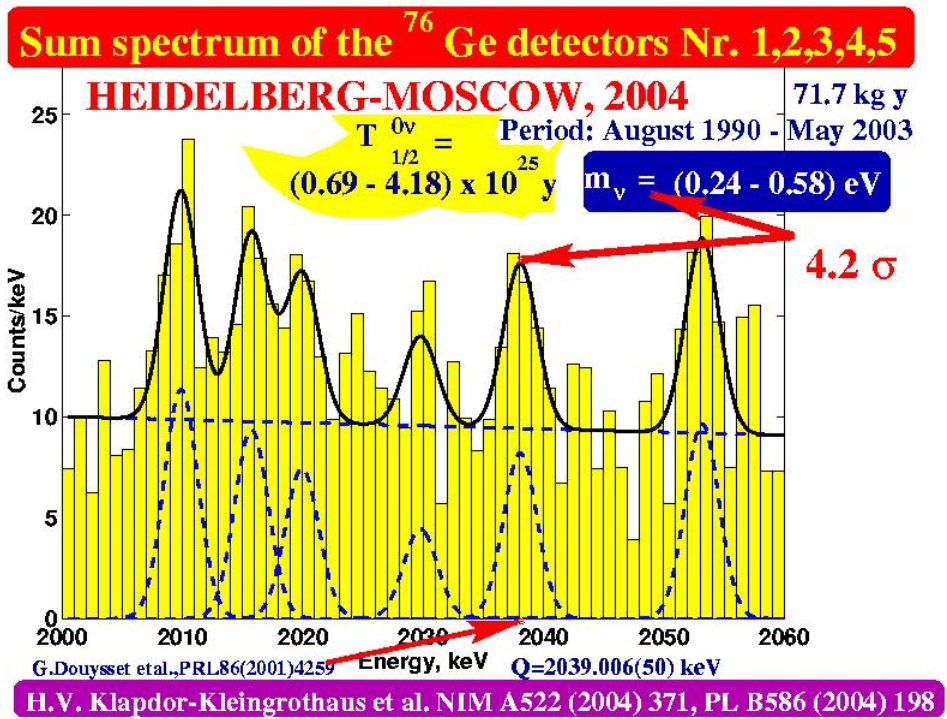

The above figure shows the sum spectrum of the $^{76}$Ge detectors 1,2,3,4 and 5 over the period August 1990 to May 2003 as recorded by the Heidelberg-Moscow experiment.

In a paper by Klapdor-Kleingrothaus, Dietz, Harney, and Krivosheina (hereafter referred to as KDHK) evidence is claimed for zero-neutrino double-beta decay in $^{76}$Ge. The high quality data, upon which this claim is based, was compiled by the 2 careful efforts of the Heidelberg-Moscow collaboration, and is well documented. However, the analysis in KDHK makes an extraordinary claim, and therefore requires very solid substantiation according to another paper "Comment on Evidence for Neutrinoless Double Beta Decay" C.E.Aalseth et al. They state that a large number of issues were not addressed in KDHK some of which are:

1. There is no null hypothesis analysis demonstrating that the data require a peak. Furthermore, no simulation has been presented to demonstrate that the analysis correctly finds true peaks or that it would find no peaks if none existed. Monte Carlo simulations of spectra containing different numbers of peaks are needed to confirm the significance of any found peaks.

2. There are three unidentified peaks in the region of analysis that have greater significance than the 2039-keV peak. There is no discussion of the origin of these peaks.

3. There is no discussion of how sensitive the conclusions are to different mathematical models. There is a previous Heidelberg-Moscow publication that gives a lower limit of $1.9 \times 10^{25}$ y (90% confidence level). This is in conflict with the "best value" of a newer KDHK paper of $1.5 \times 10^{25}$ y. This indicates a dependence of the results on the analysis model and the background evaluation.

In this paper they state that a number of other cross checks of the result should also be performed. For example, there is no discussion of how a variation of the size of the chosen analysis window affects the significance of the hypothetical peak. There is no relative peak strength analysis of all the $^{214}$Bi peaks. Quantitative evaluations should be made on the four $^{214}$Bi peaks in the region of interest. There is no statement of the net count rate of the peaks other than the 2039-keV peak. There being no presentation of the entire spectrum, is difficult to compare relative strengths of peaks. There is no discussion of the relative peak strengths before and after the single-site-event cut.

On the other hand the Heidelberg-Moscow group claims that the signal found at $Q_{\beta\beta}$ is consisting of single site events and is not a γ line. The signal does not occur in the Ge experiments not enriched in the double beta emitter $^{76}$Ge, while neighbouring background lines appear consistently in these experiments. On this basis they translated the observed numbers of events into half-lives for neutrinoless double beta decay.

The Heidelberg-Moscow experiment continued regularly from 1990 till 2003. The analysis of the full data taken with the Heidelberg-Moscow experiment in the period 2 August 1990 until 20 May 2003 is presented. The completed Heidelberg-Moscow $^{76}$Ge Experiment -71.7 kg y after 13 years of operation presents their mass calculation limit status as $m_\nu$ (eV) = 0.24 - 0.58 ( 99.997% C.L.) with the best value of 0.4 eV (95% C.L.).

## 5. The proposed MAJORANA experiment

While an unambiguous interpretation of all of the neutrino oscillation experiments is not yet possible, it is abundantly clear that neutrinos exhibit properties not included in the standard model, namely mass and flavor mixing. Accordingly, sensitive searches for neutrinoless double-beta decay (0νββ-decay) are more important than ever. Experiments with large quantities of Ge, isotopically enriched in $^{76}$Ge, have thus far proven to be the most sensitive, specifically the Heidelberg-Moscow and IGEX experiments with lower limits in half-life sensitivities $1.9 \times 10^{25}$ y and $1.6 \times 10^{25}$ y respectively. A new generation of experiments will be required to make significant improvements in sensitivity one of which is the proposed Majorana Experiment.

The Majorana Experiment is a next-generation Ge double-beta decay search which will employ 500 kg of Ge, isotopically enriched to 86% in Ge, in the form of ~200 detectors in a close-packed array for high granularity. Each crystal will be electronically segmented, with each region fitted with pulse-shape analysis electronics. A half-life sensitivity is predicted of $4.2 \times 10^{27}$ years or < $m_\nu$> ~ 0.02 - 0.07 eV, depending on the nuclear matrix elements used to interpret the data. The Majorana experiment is proposed for a US deep underground laboratory, and requires very little R&D as it stands on the technical shoulders of the IGEX experiment and other previous successful double-beta decay and low-background experiments. Furthermore, new segmented Ge detector technology has recently become commercially available, while Pacific Northwest National Laboratory (PNNL)/University of South Carolina (USC) researchers have developed new pulse-shape discrimination techniques.

Several configurations have been evaluated with respect to cryogenic performance and background reduction and rejection. It will concentrate on a conventional modular design using ultra-low background cryostat technology developed by IGEX. It will also utilize new pulse-shape discrimination hardware and software techniques developed by the Majorana collaboration and detector segmentation to reduce background. The Heidelberg-Moscow and IGEX experiments both utilized Germanium enriched to 86% in $^{76}$Ge and operated deep underground. The projection for the Majorana is that the background will be reduced by a factor of 65 over the early IGEX results prior to pulse shape analysis (from 0.2 to ~0.003 keV$^{-1}$ kg$^{-1}$ y$^{-1}$). This will occur mainly by the decay of the internal background due to cosmogenic neutron spallation reactions that produce $^{56}$Co, $^{58}$Co, $^{60}$Co, $^{65}$Zn and $^{68}$Ge in the germanium by limiting the time above ground after crystal growth, careful material selection and electroforming copper cryostats. One component of the background reduction will arise from the segmentation and granularity of the detector array.

Most of the Compton continuum consists of single Compton scatterings followed by escape of the scattered gamma ray, whereas full-energy events at typical gamma-ray energies are primarily comprised of multiple scattering sequences followed by a photoelectric absorption. The peak-to-Compton ratio can therefore be enhanced by requiring a recorded event to correspond to more than one interaction within the detector before its acceptance. In germanium detectors, this selection is usually accomplished by subdividing the detector into several segments (or providing several adjacent independent detectors) and seeking coincident pulses from two or more of the independent segments. When coincidences are found, the output from all detector segments is summed and recorded. The resulting spectrum is made up only of the full-energy peak lying above a featureless continuum that is greatly suppressed and has no abrupt Compton edges. New Ge experiments must not simply be a volume expansion of IGEX or Heidelberg–Moscow. They must have superior background rejection and better electronic stability. The summing of 200 individual energy spectra can result in serious loss of energy resolution for the overall experiment which can be avoided by segmenting n-type intrinsic Ge detectors, advanced PSD techniques and electronic stability in measurement.

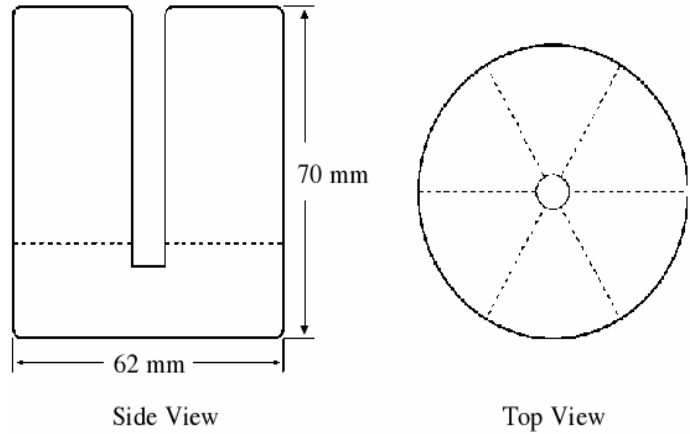

The above figure depicts a standard Ge detector segmentation scheme. This is the configuration of the SEGA detector undergoing tests by the Majorana collaboration. A configuration with six-azimuthal-segment by two-axial-segment geometry is shown in the above figure.

Efforts are thus on with the Majorana experiment for the search of neutrinoless double beta decay that would give a new shape to the standard model of physics. Majorana cannot not simply be a volume expansion of IGEX, but must have superior background rejection. As it was conclusively shown that the limiting background in at least some previous experiments has been cosmogenic activation of the germanium itself, it is necessary to mitigate those background sources. Cosmogenic activity fortunately has certain factors which discriminate it from the signal of interest. For example, while 0νββ -decay would deposit 2 MeV between two electrons in a small, perhaps 1 mm$^3$ volume, internal $^{60}$Co decay deposits about 318 keV (endpoint) in beta energy near the decaying atom, while simultaneous 1173 keV and 1332 keV gammas can deposit energy elsewhere in the crystal, most probably both in more than one location, for a total energy capable of reaching the 2039 keV region-of-interest. A similar situation exists for internal $^{68}$Ge decay. Thus deposition-location multiplicity distinguishes double-beta decay from the important long lived cosmogenics in germanium. Isotopes such as $^{56}$Co, $^{57}$Co, $^{58}$Co and $^{68}$Ge are produced at a rate of roughly 1 atom per day per kilogram on the earth's surface. Only $^{60}$Co and $^{68}$Ge have both the energy and half-life to be of concern. To pursue the multiplicity parameter, firstly, the detector current pulse shape carries with it the record of energy deposition along the electric field lines in the crystal; that is, the radial dimension of cylindrical detectors. This information may be exploited through pulse-shape discrimination. Secondly, the electrical contacts of the detector may be divided to produce independent regions of charge collection. The ability of new techniques to be easily calibrated for individual detectors makes them practical for large detector arrays. Calibration for single-site event pulses was trivially accomplished by collecting pulses from thorium ore; the 2614.47-keV gamma ray from $^{208}$Tl produces a largely single-site double-escape peak at 1592.47 keV. The PSD discriminator was then calibrated to the properties of the double-escape peak A slightly improved double-escape peak was be made from the $^{26}$Al gamma ray of 2938.22-keV. The double-escape appears at 1916.22 keV, only about 120 keV away from the expected region of interest for 0νββ-decay. The obvious and direct use of pulse- shape discrimination and segmentation is the rejection of cosmogenic pulses in the germanium itself. However, the approach should be also effective on gamma rays from the shielding and structural materials. The background effects of neutrons of both high energy (cosmic muon generated) and low energy (fission and (α,n) from rock) could be protected by the segmentation and granularity of the detectors. These neutrons could also produce other unwanted activities like the formation of $^3$H and $^{14}$C in nitrogen from high and low energy neutrons, respectively. Fortunately, Majorana detectors will not be surrounded by nitrogen at high density.

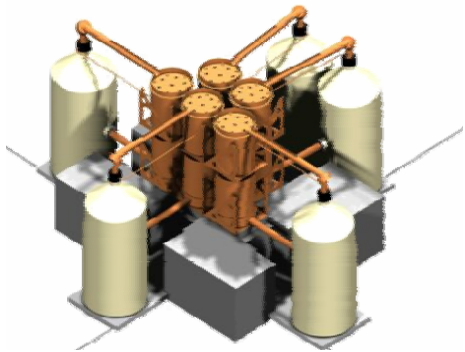
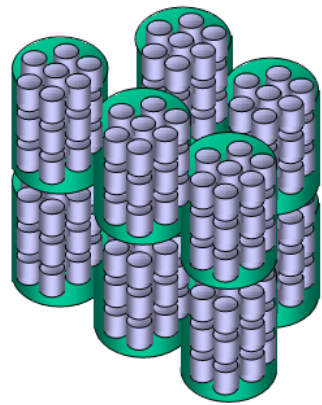

The GERDA (GERmanium Detector Assembly), which is another next generation $^{76}$Ge double beta decay experiment at the Gran Sasso Underground Laboratory, has projected a sensitivity in the half-life of the 0νββ-decay mode which is less than the proposed Majorana experiment. In conclusion, the Majorana project has been designed in a compact, modular way such that it can be built and operated with high confidence in the approach and the technology. The initial years of construction will allow alternate cooling methods to be employed if they have an advantage and should they be shown to overcome long-term concerns due to surface contamination, muon-induced ions, and diffusion. The Majorana Collaboration has made an extensive analysis of the predicted backgrounds and their impact on the final sensitivity of the experiment. The Majorana experiment represents a great increase in Ge mass over IGEX with new segmented Ge detectors and the newest electronic systems for pulse-shape discrimination. Their conclusion is that with 500 kg of Ge, enriched to 86% in the isotope $^{76}$Ge, the Majorana array operating over 10 years including construction time, can reach a lower limit on $T_{1/2}^{0\nu}$ of $4\times10^{27}$ years. This corresponds to an upper bound of $<m_\nu>$ of $0.038 \pm 0.007$eV. One advantage of $^{76}$Ge is that it may well be a candidate for a future more reliable microscopic calculation of the 0ν ββ- decay nuclear matrix element.

## 6  Conclusion

Neutrinoless double beta decay is thus one of the most sensitive approaches with great perspectives to test particle physics beyond the Standard Model. The possibilities to use 0νββ decay for constraining neutrino masses, left–right symmetric models, SUSY and leptoquark scenarios, as well as effective lepton number violating couplings, have been reviewed. It is a very sensitive probe to the lepton number violating terms in the Lagrangian such as the Majorana mass of the light neutrinos, right-handed weak couplings involving heavy Majorana neutrinos, as well as Higgs and other interactions involving violation of chirality conservation.

In search for neutrinoless double beta decay $^{76}$Ge as the source material has multiple advantages. It has high resolution (< 4 keV at $Q_{\beta\beta}$) with no background from 2ν mode. A huge leap in sensitivity is possible applying ultra-low background techniques and 0ν- ββ signal discrimination. There can be a phased approach in the experiment with the increment of target mass. The source and detector are the same material thereby reducing background and maintaining the 4π geometry and the only way to scrutinize 0ν – DBD claim on short time scale: since it tests $T_{1/2}$ and not $m_\nu$. The consequences of Neutrinoless Double Beta Decay are- [1] Total Lepton number violation: The most important consequence of the observation of neutrinoless double beta decay is that lepton number is not conserved. This is fundamental for particle physics. [2] Majorana nature of neutrino: Another fundamental consequence is that the neutrino is a Majorana particle. Both of these conclusions are independent of any discussion of nuclear matrix elements. [3] Effective neutrino mass: The matrix element enters when we derive a value for the effective neutrino mass - making the most natural assumption that the 0νββ decay amplitude is dominated by exchange of a massive Majorana neutrino.

**Acknowledgements**
I would like to thank the IGEX collaboration, the Heidelberg-Moscow collaboration and the Majorana collaboration for having used information from their experimental works to write up this brief review.